\documentclass[showpacs, twocolumn, aps, prb, longbibliography, floatfix]{revtex4-2} 

        \expandafter\let\csname equation*\endcsname\relax
        \expandafter\let\csname endequation*\endcsname\relax
        \usepackage{amsmath,bm}
        \usepackage{multirow,booktabs}   
        \usepackage{amsthm}
        \usepackage{mathrsfs}
        \usepackage{amssymb}
        \usepackage{longtable}   
        \usepackage{amsfonts} 
        \usepackage{graphicx}
        \usepackage{url}
        \usepackage{hyperref}
        \usepackage[columnwise]{lineno}  
        \usepackage{color}
        \usepackage{comment} 
        \usepackage{epstopdf}  
        \renewcommand{\comment}[1]{}
    
            \makeatletter
            \newsavebox{\@brx}
            \newcommand{\llangle}[1][]{\savebox{\@brx}{\(\m@th{#1\langle}\)}%
              \mathopen{\copy\@brx\kern-0.5\wd\@brx\usebox{\@brx}}}
            \newcommand{\rrangle}[1][]{\savebox{\@brx}{\(\m@th{#1\rangle}\)}%
              \mathclose{\copy\@brx\kern-0.5\wd\@brx\usebox{\@brx}}}
            \makeatother

        \newcommand{\vgm}{$v_\text{g,max}$ }

        \usepackage{bibentry}

        \newcommand{\ignore}[1]{}
        \newcommand{\bibfnamefont}[1]{#1}
        \newcommand{\bibnamefont}[1]{#1}

\mathchardef\mhyphen="2D

\begin{document}


        
        \title{Slow transport and bound states for spinless fermions with long-range Coulomb interactions on one-dimensional lattices}

    \author{Zhi-Hua Li}
    \address{School of Science, Xi'an Technological University, Xi'an 710021, China}

    \begin{abstract}
            We study  transport and relaxation of spinless fermions with
            long-range Coulomb  interactions at high temperatures through numerical
            simulations of out-of-equilibrium dynamics.  We find that the
            transport and relaxation are continuously slowing down for
            increasing coupling $V$, and that there is a transition in the type of
            transport.  For intermediate couplings, the system exhibits normal
            diffusive transport but the time scale for the onset of that is
            long. For large couplings, it exhibits subdiffusive transport, while
            at the same time the relaxation time diverges exponentially with
            system lengths, featuring an MBL-like phase.  We attribute the slow
            transport to formation of slow bound states and stable clusters of particles. For
            few-particle systems we prove existence, visualize the slowness and
            analyze collision properties of the bound states. For many particles
            at high densities there should be a hierarchy of clusters of
            particles  on many different length scales. We argue that at large
            couplings the average maximal size of the stable clusters should
            scale linearly with the length of the lattice, which is in
            accordance with the MBL-like behavior. 
            %
            %
    \end{abstract} 


    \maketitle
    \modulolinenumbers[5]

\section{introduction}

    Understanding how macroscopic hydrodynamics emerges from microscopic laws is
    an important question, which is generally too difficult to be tractable.
    However, in recent years a breakthrough has been made for integrable
    systems, which is coined generalized hydrodynamics
    (GHD)\cite{bertini2016transport,castroalvaredo2016emergent}. It has been
    established in GHD that integrable systems can support ballistic transport at
    finite temperatures due to existence of infinite many conserved charges.
    Various transport quantities for many 1D quantum integrable models have been
    calculated \cite{bertini2020finite}. In particular, it has been applied to 
    the XXZ model for its ballistic\cite{ilievski2017microscopic},
    diffusion \cite{nardis2019diffusion} and
    superdiffusion \cite{ljubotina2019kardar,gopalakrishnan2019kinetic} regimes.

    Although GHD is successful for integrable systems, integrability is rare in
    the real world and there are always various perturbations to break it.  On
    the one hand, some groups have attempted to incorporate (weak) integrability
    breaking terms into GHD \cite{friedman2020diffusive,
    bastianello2021hydrodynamics}, since one can always use the Bethe ansatz 
    vectors as a base for generic models, being them integrable or not. 
    On the other hand, transport of many non-integrable models has been studied
    numerically. These include the XXZ model with dimerization and frustration
    \cite{langer2009real}, staggered field \cite{bulchandani2020superdiffusive},
    and spin ladders \cite{zotos2004high,jung2006transport}, just to name a few.
    Although in the majority cases transport becomes diffusive, as it is
    expected, there are other cases where transport is anomalous
    \cite{bulchandani2020superdiffusive,nardis2019diffusion,chen2023superdiffusive}.
    Our understanding of transport of non-integrable models is still far from
    complete.

    In the previous numerical studies, the integrability breaking terms are
    mostly short-ranged, and the transport quantities are extracted from
    dynamics in the linear response regime for relatively short time scales, so
    that usually normal diffusive transport is found.  In this paper we study
    transport and relaxation of a 1D fermion model with translation invariant
    long-range interactions through numerical simulations of 
    out-of-equilibrium dynamics, for longer time scales and a wide range of
    coupling strengths. We found that the transport and relaxation are 
    continuously slowing down for increasing coupling $V$, and there is even a
    transition of the type of transport: For intermediate couplings, the
    relevant quantities in the dynamics would attain the values signaling normal
    diffusive transport or thermalization, but the processes for reaching those
    values are logarithmically slow in time.  For large couplings, the system
    displays subdiffusive transport, and at the same time, the relaxation time
    diverges exponentially with the system sizes, showing a lack of
    thermalization in the thermodynamic limit. 
    
    
    To understand the slow transport we studied certain few-body problems of the
    model. We find that there are various $n$-particle bound states because  of
    the limited band width ($\sim\lambda$) of the lattice model as well as the
    long-range interactions. And the group velocities of the bound states can be
    exponentially slow in $n$ when $V\gtrsim \lambda$. 
    Then, for many particles with high densities and  at large
    couplings, there should be slow moving clusters of particles on different length
    scales. The exponential divergence of the relaxation time is explained by
    possible giant immobile clusters, whose sizes may be proportional to the
    length of the lattice.
    
    Several works have already discovered divergence of relaxation times in
    certain disorder-free models \cite{kagan1984localization,
    roeck2014asymptotic, grover2014quantum, schiulaz2014ideal,
    schiulaz2015dynamics,yao2016quasi,mondaini2017many,bols2018asymptotic,michailidis2018slow,spielman2022slow},
    which was dubbed quasi-many body localization (MBL) states
    \cite{yao2016quasi} or asymptotic localization
    \cite{roeck2014asymptotic,bols2018asymptotic}.
    Some of these works manually introduced two components of fast and slow
    particles to realize such states.  While we take the above point of view
    that there can be self-generated slow bound states or clusters
    \cite{kagan1984localization}, which is more natural and closer to realizable
    physical systems such as carbon nanotube or cold atom systems
    \cite{baumann2010dicke,landig2016quantum}.  
    Besides, there has been ambiguity about the nature of the quasi-MBL states
    \cite{papic2015many}.  We elucidate that the quasi-MBL states can be
    coincided with subdiffusion transport, and that a possible structure of the
    quasi-MBL states could be a hierarchy of stable clusters of particles but
    with internal resonant dynamics. Since slow bound states widely exist in
    quantum lattice models, we expect that they should play an important role in
    formulating a general theory of transport for these models, especially in
    the large coupling regime. 
    
    
    The rest of the paper is organized as follows: Sec. \ref{sec:model}
    introduces the model Hamiltonian and observables.  Sec. \ref{sec:results}
    presents numerical results demonstrating the slow transport and relaxation
    properties. Sec. \ref{sec:bound:state} delivers a systematic study of the
    bound states of the model, including properties of their spectra, group
    velocities and scatterings, based on which the transport and relaxation
    processes are interpreted.  Finally, conclusions are drawn in
    Sec. \ref{sec:conclusion}.

\section{model and observables}        \label{sec:model}
    The model considered here consists of a chain with $L$ sites, 
    \begin{equation}
        \begin{aligned}
        \hat H = & -\lambda \sum\limits_i { (\hat c_i^ \dag {{\hat c}_{i + 1}} + h.c.)}
        \\
         & + \sum\limits_{i < j}{ 
        {\frac{V}{|j-i|^\alpha}({{\hat n}_{i}} - 1/2)({{\hat n}_{j}} - 1/2)} } 
        \label{eq:ham}
        \end{aligned}
        \end{equation}
    where $\hat c^\dagger_i$ ($\hat c_i$) is creation (annihilation) operator of
    a spinless fermion, and $\hat n_i$ is a fermion density operator at site
    $i$.  The interactions decay in power laws governed by an exponent $\alpha$.
    This model can be rewritten via the Jordan-Wigner transformation as a
    quantum spin model $\hat{H} =  - \frac{J}{2}\sum_i {(\hat S_i^x\hat
    S_{i + 1}^x + \hat S_i^y\hat S_{i + 1}^y)}  + \Delta \sum_{i < j} {|j
    - i{|^{ - \alpha }}\hat S_i^z\hat S_j^z} $,  after the identification of
    $J=2\lambda$ and $V=\Delta$. 
    In particular, at $\alpha=\infty$, it reduces to the  XXZ model (since the
    sign of $J$ is unimportant). 
    By virtue of this, the languages for fermion and spin systems will be used
    interchangeably, and charge transport can be rephrased as spin transport.
    In the language of spins, the total magnetization $ {\hat
    S^z_\text{tot}}=\sum_{i} {\hat S^z_i}$ is conserved. When there is 
    an inhomogeneity in spin densities, spins are transported, which can be
    quantified by measuring the spin current operator ${\hat j_i} = J(\hat S_i^x
    \hat S_{i + 1}^y- \hat S_i^y \hat S_{i + 1}^x)$.  

     The ground state properties of this model have been studied in Refs.
     \cite{capponi2000effects,hohenadler2012interaction,li2019ground,ren2020entanglement}.
     Here we investigate its transport and relaxation dynamics at high
     temperatures. We fix $\alpha=1$ if not otherwise specified, which
     corresponds to the unscreened Coulomb potential.  And we focus on the range
     of $V \gtrsim 2$, where as we will see, the transport is slow due to
     formation of slow bound states. The unit $\lambda=\hbar=1$ is used, which
     also sets the unit of time to be $\hbar/\lambda=1$. Next we introduce two
     quantities to characterize transport and relaxation of the model,  both of
     which are extracted from out-of-equilibrium dynamics. 

    The first quantity is a transport exponent extracted from a bipartite quench dynamics.  
    The initial state of the dynamics is a mixed-type domain-wall state 
    \cite{ljubotina2017spin},  
    \begin{equation}
        \rho(t=0) \propto \left(1+\mu \sigma^{z}\right)^{\otimes \frac{L}{2}}
        \otimes\left(1-\mu \sigma^{z}\right)^{\otimes \frac{L}{2}},
        \label{eq:rho}
        \end{equation}
    where $\mu$ induces an initial imbalance of magnetization between the left and right
    halves of the chain:  
    $\langle \hat S^z_{i\leq L/2, i>L/2} \rangle=\pm\tfrac{1}{2}\mu$. When
    $\mu=0$, the system is in the maximally mixed state, corresponding to infinite
    temperature. 
    A small $\mu$ will be used, which implies that the system is weakly-polarized
    and at a high temperature.  The time evolved density matrix is given by 
    $\rho(t)=e^{-i\hat{H}t} \rho(0) e^{i\hat{H}t}$ as usual, which can be solved
    numerically by using e.g. matrix product state (MPS) based algorithms (see below). 
    
    Once $\rho(t)$ is obtained, one can characterize the transport properties by
    the evolution of magnetization $m(i,t)=tr[\rho(t) \hat S^z_i]/tr[\rho(t)]$
    and by the current $j_i(t)=tr[\rho(t) \hat j_i]/tr[\rho(t)]$.  It is
    expected that, at \emph{large} time, the magnetization will have a scaling
    form $m(i, t)=\varphi(\xi)$, with the scaling variable $\xi=(i-L/2)/t^z$,
    and that the current across the center cut should behave as $j_{{L/}{2}}
    \sim t^{z-1}$.  Then the type of transport can be classified by the
    dynamical exponent $z$: it is ballistic if $z=1$, diffusive if $z=0.5$ and subdiffusive if
    $z<0.5$.  In practice, $z$ is time dependent before reaching its asymptotic
    value, which may provide extra valuable information about the dynamics.  A
    convenient way to extract the time dependent transport exponent is first to
    calculate the accumulation of spins transported through the center cut of
    the chain  
    \begin{align}
        \label{eq:delta:m}
        \Delta M(t) 
        & = \sum_{i=1}^{L/2}{ [ \frac{\mu}{2} - m(i,t)  ]}  = \sum_{i=L/2+1}^{L}{ [m(i,t) + \frac{\mu}{2}]}   \nonumber \\ 
        & = \int_{0}^{t} j_{\tfrac{L}{2}}\left( t^{\prime}\right) \mathrm{d}
        t^{\prime} \propto t^{z}, 
    \end{align}
    and then to take a logarithmic derivative, that is 
    \begin{equation}
        \label{eq:z:def}
        z(t)=d\ln(\Delta M)/d \ln(t). 
    \end{equation}
   
    The second quantity is extracted from the relaxation of spatial inhomogeneities of
    particle densities, which can be used to probe possible localized phases. 
    Specifically, starting from an initial state $|\psi(0)\rangle$ which is a random
    classical state, such as $|01001\cdots 010\rangle$,  its relaxation process
    can be measured by  \cite{schiulaz2015dynamics}
    \begin{equation} \label{eq:inhomo} \Delta {\rho^2_{\psi}}(t) =
        \frac{1}{L}\sum\limits_{i = 1}^L {{{\left[ \langle \psi(t) | {{{\hat
        n}_{i + 1}}(t) - {{\hat n}_i}(t)} | \psi(t) \rangle \right]}^2}},
    \end{equation} where $|\psi(t)\rangle =e^{-i\hat H t}|\psi(0)\rangle$ is the
    time evolved state. 
    Since we are interested in the dynamics at infinite temperature, an average
    value $\langle \Delta \rho^2_\psi(t) \rangle $ is taken for $|\psi(0)\rangle$
    drawn from a sector with a fixed filling factor $\nu$ (number of particles
    divided by $L$).  After normalizing it with its initial value, one arrives
    at 
    \begin{equation}
        f(t)\equiv \frac{\langle \Delta \rho^2_\psi(t) \rangle}{\langle \Delta
        \rho^2_\psi(0)
        \rangle}. 
        \label{eq:f:def}
        \end{equation}
    Then one asserts that the system is localized if $f$ remains finite for
    infinite time, otherwise, it thermalizes. 
    
    Based on the time dependences of the two quantities $z$ and $f$, we also
    define two important time scales for each of them: a time scale $\tau$ for
    when $z$ reaches 0.5, signaling diffusive transport, and a relaxation time
    $\tau_1$ for when $f$ reaches 0, provided they do reach these values.  
    

       

\section{numerical results}  \label{sec:results}
        
\subsection{transport exponent $z(t)$} \label{sec:results:z}

    
    We use a two-site version of the MPS-based time dependent variation principle
    algorithm (TDVP) \cite{haegeman2014unifying}  to simulate the time evolved
    density matrix $\rho(t)$. This algorithm can deal with Hamiltonians with
    long-range interactions through a matrix product operator (MPO) technique
    \cite{crosswhite2008applying}. The parameter $\mu$ for the initial $\rho(0)$
    is set to be 0.01. The density matrix $\rho(t)$  is evaluated for $t$ up to
    1000.  The system size $L$ ranges from 128 to 384, depending on the
    coupling $V$. Larger $L$ is needed for  smaller $V$, to avoid boundary
    effects. The largest bond dimension of the MPS used is 320. 
    
\begin{figure}   
          \centering
          \scalebox{0.55}[0.55]{\includegraphics{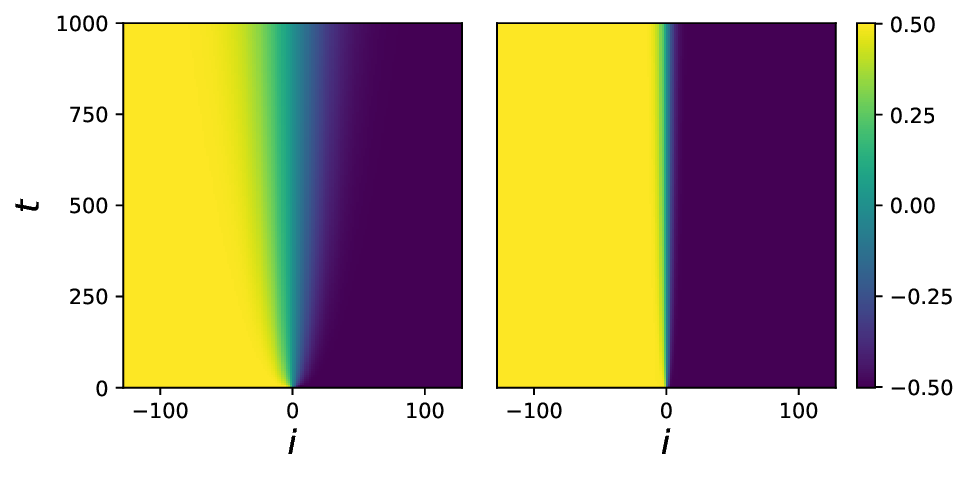}}
          \caption{\label{fig:sz:x:t} 
          Time evolution of the spin density $\mu^{-1} m(i,t)$ at $V=4$
          (left panel) and $V=16$ (right panel) on a chain with $L=256$ sites.
          The origins of the $i$-axes are shifted to $L/2$.  
          }
          \end{figure}

    We first show the evolution of the magnetization $m(i,t)$ for two couplings
    $V=4$ and $16$ in Fig. \ref{fig:sz:x:t}. Slowing down of transport with
    increasing $V$ can be intuitively  seen from this figure. It is also due to this
    fact that we can simulate the quench dynamics for relatively long times   
    with  moderate costs. It is cumbersome to extract the transport exponent $z$
    from the scaling form of $m(i,t)$, and even more difficult to obtain its time
    dependence in this way. So we extract the time dependent exponent $z(t)$ using
    Eqs.  \eqref{eq:delta:m} and \eqref{eq:z:def}, instead. 
    

    
    \begin{figure}   
          \centering
          \scalebox{0.55}[0.55]{\includegraphics{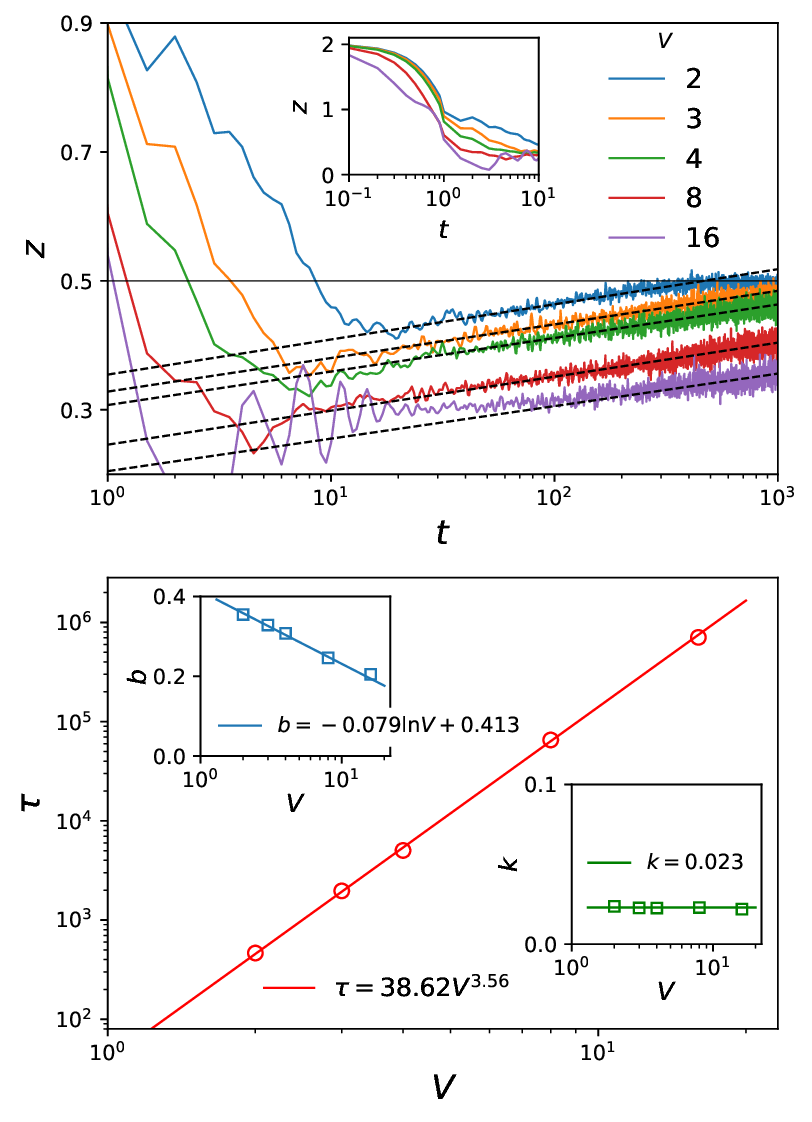}}
          \caption{\label{fig:z:t} 
          Upper panel: dependence of the dynamical exponent $z$ on time for
          different coupling strengths. The inset panel shows the details at early
          times.  The $z$ values are expected to reach 0.5 (the horizontal solid
          line) at large times. The dashed lines are fittings to Eq.  
          \eqref{eq:z:log:t}.  Lower panel: the time scale $\tau$ (and the
          fitting parameters $b$ and $k$ in the two inset panels) as a function
          of $V$. The values of $\tau$ are determined through Eq.  
          \eqref{eq:tau}. Symbols are data, while the solid lines are fitting
          functions as indicated in the legends. 
        } 
    \end{figure}
    

    
    The upper panel of Fig. \ref{fig:z:t} shows $z(t)$ for several intermediate
    coupling strengths.  For each $V$ there are multiple stages in the dynamics:
    (i) $z$ drops from a super-ballistic value around 2 at $t\approx 0$ to
    around the ballistic value of 1 at $t=1$ (see the inset panel). (ii) $z$
    continues dropping for $t>1$, then reaches a minimum value, and then it may
    fluctuate until $t=10\sim 100$, which depends on $V$. This is a transient
    period connecting (i) and the next stage. (iii) $z$ increases very slowly
    with time which can be  fitted approximately by  a logarithmic function
    \begin{equation} 
        \label{eq:z:log:t} z(t)= k \ln(t)+b, 
    \end{equation} 
    with two fitting parameters $k$ and $b$. This logarithmic process terminates
    when $z$ reaches 0.5. After that the system enters a steady state i.e.  stage
    (iv). This final stage is clearly seen only for $V=2$, due to restrictions in
    the simulation time. But we expect that the transport should become
    diffusive for other values of $V$ in the figure, although the time
    scale $\tau$ for that to happen is much longer for larger $V$.
    
   
     It is worthwhile to obtain a quantitative relationship between $\tau$ and
     $V$.  Then a quantified value of $\tau$ is needed. To this end we use Eq.  
    \eqref{eq:z:log:t} and the fitting parameters $k$ and $b$ to obtain an estimated value
    of $\tau$, namely, by solving $z(\tau)=0.5$, which yields
    \begin{equation}
        \label{eq:tau}
        \tau=e^{(0.5-b)/k}. 
    \end{equation}
    The result is shown in the lower panel of Fig. \ref{fig:z:t}. 
    It turns out that the estimated value of
    $\tau$ scales with  $V$ in a power law  
    \begin{equation}
        \label{eq:tau:vs:V} 
        \tau\propto V^\kappa, 
    \end{equation}
    with an exponent $\kappa\approx 3.56$. Since the dependence of $\tau$ on $V$ comes from that of
    $k$ and $b$ on $V$, it is beneficial to also look at the latter ones. One can
    see in the two inset panels that,   $b$ decreases with $V$ which can be
    fitted in the form $b=A\ln V + B$, while $k$ seems to be a constant, these
    leading to a refined form of Eq. \eqref{eq:z:log:t}, 
    \begin{equation}
        \label{eq:z:log:t:refined} 
        z(t)=k\ln(t) + A\ln V + B, 
    \end{equation}
    with the  constant coefficients $k=0.023$, $A=-0.079$ and $B=0.413$.

    \begin{figure}  
          \centering
          \scalebox{0.55}[0.55]{\includegraphics{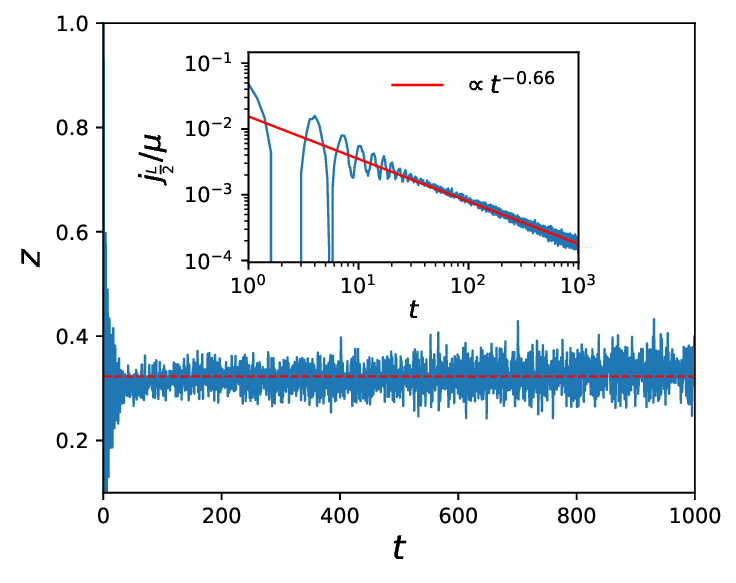}}
          \caption{\label{fig:z:t:large:V} 
          Time dependence of the dynamical exponent $z$ for $V=32$. The exponent
          oscillates around $z=0.32$ (the red dashed line) at late times. Inset panel
          shows the current flowing through the center cut, which can be fitted by a 
          power function (the red solid line)  at late times.  
          }
          \end{figure}
    In the above we have shown that, for intermediate couplings, the system
    should enter a steady state with normal diffusive transport, only that the
    time scale $\tau$ for the onset of it can be very long. In fact, for large
    couplings, the system may never reach diffusion and Eqs. \eqref{eq:z:log:t}
    and \eqref{eq:tau:vs:V} are no longer valid. We illustrate this in
    Fig. \ref{fig:z:t:large:V} for the coupling $V=32$. One can see that $z$
    keeps oscillating around a constant value at late times  that is below 0.5
    (the oscillation may come from numerical errors when taking the logarithmic
    derivative in Eq. \eqref{eq:z:def}). This indicates that the transport is
    further slowed down at large $V$ and a dynamical phase transition to subdiffusion occurs. 
    
    Note that the above quantities drawn from the bipartite quench dynamics are
    essentially the thermodynamic limit results. In practice we find that it
    is harder to simulate the dynamics for even larger $V$ using the TDVP
    algorithm.  Next we study the other quantity $f(t)$ for short finite
    systems, but for much larger couplings and longer time scales.



\subsection{relaxation quantity $f(t)$}  \label{sec:results:f}
    

    We use an exact diagonalization (ED) algorithm \cite{weinberg2017quspin} to
    calculate the time evolution problem $|\psi(t)\rangle =e^{-i\hat H
    t}|\psi(0)\rangle$, where periodic boundary conditions (PBC) are used.  Each
    data of $f$ shown below are obtained by using 300 realizations of
    $|\psi(0)\rangle$ in the sector of $\nu=\tfrac{1}{2}$. 

    \begin{figure}   
          \centering
          \scalebox{0.55}[0.55]{\includegraphics{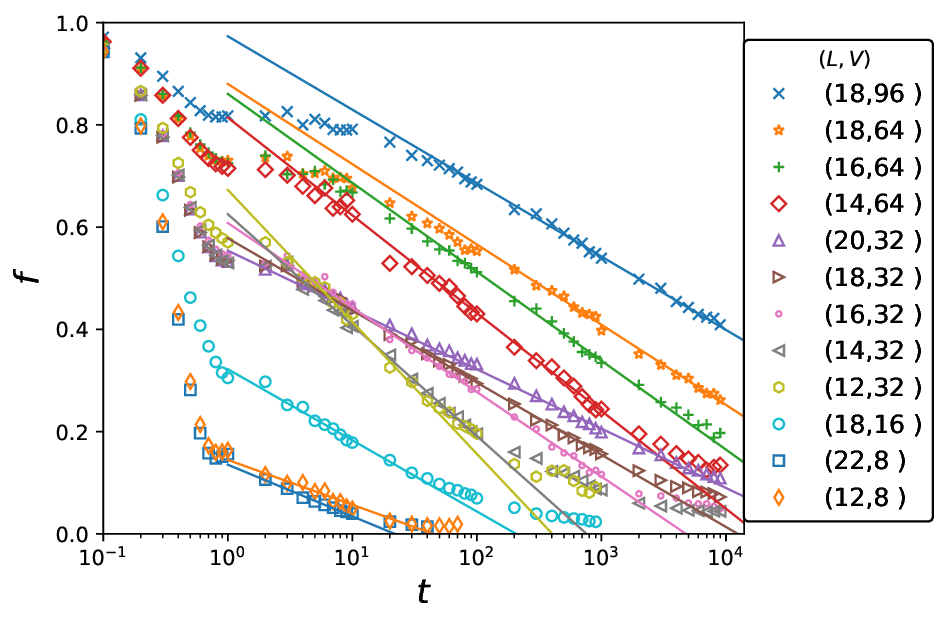}}
          \caption{\label{fig:inhomogenity} 
          Relaxation of spatial density inhomogeneity for several pairs of 
          coupling strengths and system sizes. Solid lines are fittings to 
          Eq. \eqref{eq:f:log:t}. 
          }
          \end{figure}
    
    Fig. \ref{fig:inhomogenity} shows relaxation of the inhomogeneity $f$ for
    several $(L, V)$ pairs with $L\in[12,22]$ and $V\in[8,96]$, and for $t$ up
    to $10^4$.  For each $(L,V)$ pair, there are multiple stages in the
    dynamics: (i) For $t\in [0,1]$, $f$ decays fast, whose rate depends mainly
    on $V$ but barely on $L$. (ii) A transient period connects (i) and the
    next stage. This period lasts till $t\sim O(1)$ for smaller $V$, and longer
    till $t\sim O(10)$ for larger $V$. (iii) A slow approximately logarithmic
    decay, which can be
    fitted by 
    \begin{equation}
        \label{eq:f:log:t}
        f(t) =-k_1\ln t +b_1,   
    \end{equation}
    with two fitting parameters $k_1$ and $b_1$. The decay rate depends mainly
    on $L$ but only slightly on $V$. This stage terminates, when $f$ has dropped
    to a low level, for example to $f\approx 0.1$ at $t\approx 80$ for
    $(L,V)=(18, 16)$. Then the final stage (stage (iv)) starts, during which $f$
    decays even slower and finally approaches zero.  The final stage is only
    visible for small $L$ and $V$ in the figure due to limitations in the time
    of the simulations, but we assume that there is still such a stage for other
    cases. That is to say the system is expected to thermalize for all finite
    $L$ and $V$.  
    
    Since qualitatively the relaxation time $\tau_1$ for $f$ approaching 0
    increases with larger $L$, an intriguing question is then: Would
    the relaxation time diverge in the thermodynamic limit? Then a quantitative
    value of $\tau_1$ is needed.  To this end, we utilize the fitting function
    Eq. \eqref{eq:f:log:t} of stage (iii) to obtain an estimated value of $\tau_1$
    (or a lower bound of it). Namely,  for each $(L,V)$ pair, it is
    determined by the fitting parameters, 
    \begin{equation} 
        \label{eq:tau1} \tau_1=e^{b_1/k_1}.
        \end{equation}
    Then we study how this estimated relaxation time changes with $L$ and
    $V$.

\begin{figure}
          \centering
          \scalebox{0.55}[0.55]{\includegraphics{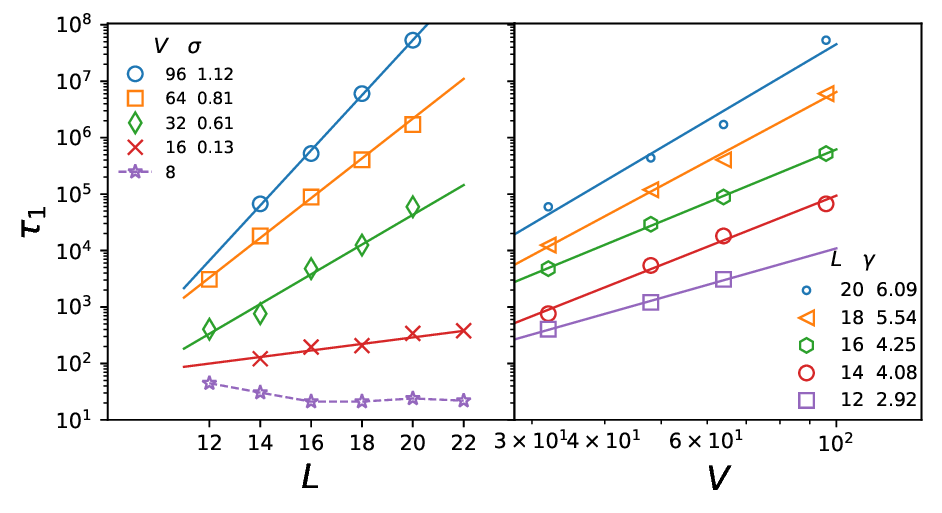}}
          \caption{\label{fig:tau1:vs:L} 
          Left panel: dependence of relaxation time on the system size, with fixed
          couplings. Solid lines are fittings to  Eq. \eqref{eq:tau1:L}, with
          the exponents $\sigma$ shown in the legend.  The dashed line is a guide for
          the eye.  
          Right panel: dependence of relaxation time on the coupling strength, with
          fixed lattice lengths. Lines are fittings to Eq. \eqref{eq:tau1:V}, with
          the exponents $\gamma$ shown in the legend.
          }
          \end{figure}

    The left panel of Fig. \ref{fig:tau1:vs:L}  shows dependence of $\tau_1$ on $L$
    with fixed couplings.  For small $V$, $\tau_1$ saturates with increasing
    $L$, which means that the system thermalizes in thermodynamic limit. Whereas, for
    each large $V$, $\tau_1$ grows exponentially with $L$,
    \begin{equation}
        \label{eq:tau1:L}
        \tau_1 \propto e^{\sigma L}, 
    \end{equation}
    with an exponent $\sigma$ possibly depending on $V$.  This relation means
    lack of thermalization and corresponds to a quasi-MBL phase introduced in
    Ref.  \cite{yao2016quasi}.  So there is a transition between the small and
    large coupling regimes. However, we are not meant to locate a transition
    point $V_c$  precisely in this paper. Next the dependence of $\tau_1$ on $V$
    for each lattice size in the large coupling regime is shown in the right
    panel.  For each $L$, $\tau_1$ grows with $V$ in a power law, 
    \begin{equation}
        \label{eq:tau1:V}
        \tau_1 \propto  V^\gamma,    
    \end{equation}
    with an exponent $\gamma$ possibly depending on $L$. 
    \begin{figure}
          \centering
          \scalebox{0.55}[0.55]{\includegraphics{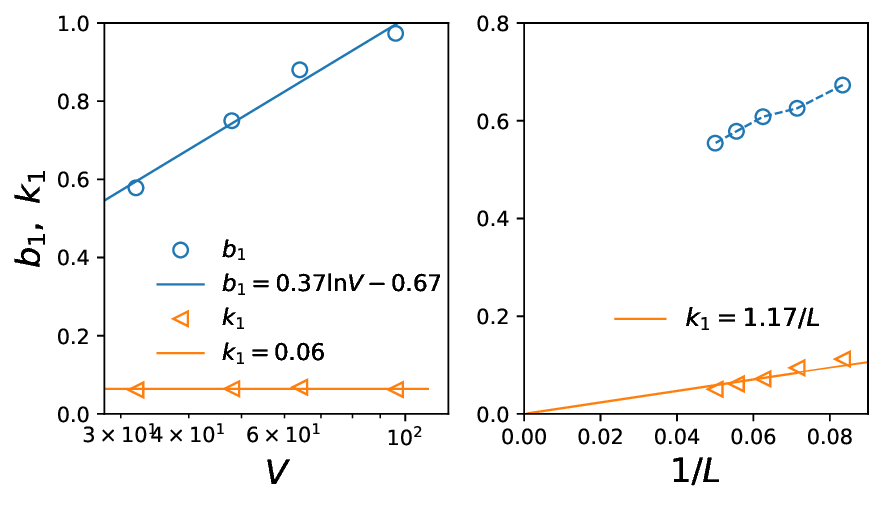}}
          \caption{\label{fig:bk:vs:lv}
          Dependences of the fitting parameters $b_1$ (symbols of circles) and $k_1$
          (symbols of triangles)  on the coupling strength $V$, for fixed $L=18$ (left
          panel); and dependences of them on the inverse system size $1/L$, at 
          fixed $V=32$ (right panel). Solid lines represent fitting functions, as
          indicated in the legend; the dashed line is a guide for the eye. 
          }
          \end{figure}
    
    It is tempting to obtain a full function relation $\tau_1(L,V)$ in the
    regime $V>V_c$.  In fact, since $\tau_1$ is determined by the two parameters
    $b_1$ and $k_1$, that can be partially  achieved by studying how the two parameters 
    depend on $L$ and $V$.  First we fix an $L$, say $L=18$, and look at how 
    they depend on $V$.  One can see from the left panel of
    Fig. \ref{fig:bk:vs:lv} that $k_1$ obviously does not depend on $V$, while
    $b_1$ increases logarithmically with $V$. The latter can be fitted in the
    form $b_1=C \ln V+D$, where the coefficients $C$ and $D$ may be
    $L$-dependent. Next we fix a $V$, say $V=32$, and look at how they depend
    on $L$. One can see from the right panel that $k_1$ is proportional to
    the inverse system size, as $k_1=u/L$, where the coefficient $u\approx 1.17$
    (note that $u$ should be a constant, not depending on $V$). From these, we
    obtain a refined form of Eq. \eqref{eq:f:log:t} 
    \begin{equation}
        \label{eq:f:refined}
        f(t)= - \frac{u}{L}\ln(t) + C\ln V +D, 
    \end{equation}
    and then 
     \begin{equation}
        \label{eq:tau1:LV}
        \tau_1(L,V)=e^{(C\ln V +D)L/u}
    \end{equation}
    for the regime of $V>V_c$. 
        
\begin{figure}
          \centering
          \scalebox{0.55}[0.55]{\includegraphics{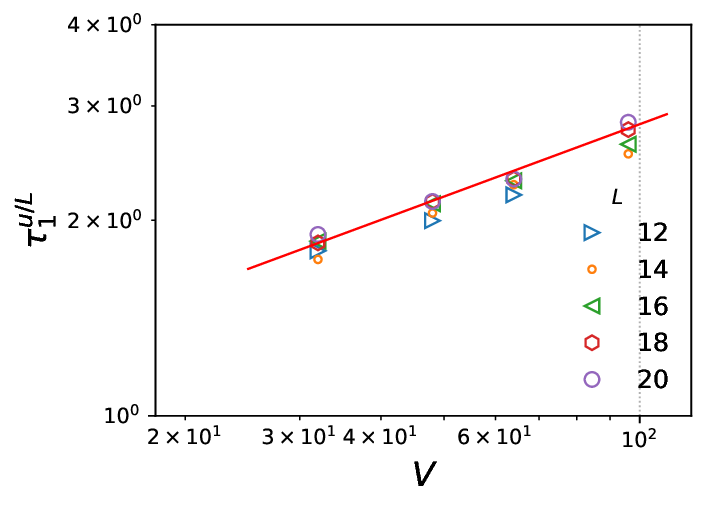}}
          \caption{\label{fig:data:collapse} 
          Rescaled relaxation time $\tau_1^{u/L}$ vs. the coupling strength for
          different system sizes.  The solid line represents the function
          $\tau_1^{u/L} = e^{C\ln V +D}$, with $u=1.17$, $C=0.37$ and $D=-0.67$.
          The near-collapse of data to this line for each $L$ validates this
          function. 
          }
          \end{figure}

    The problem remaining is to determine the function relations $C(L)$ and
    $D(L)$. In fact, to make Eq. \eqref{eq:tau1:LV} consistent with Eq. 
    \eqref{eq:tau1:L}, the only possibility is $C(L)= C_{L=\infty} +E/L +
    \mathcal{O}(1/L^2)$ ($E$ being a coefficient); $D(L)$ should have a similar
    form for the same reason. Therefore $C$ and $D$ can be taken as constants
    for large $L$. 
    We also note that, when $C\ln(V)+D=0$, $\tau_1$ will be finite in the
    thermodynamic limit. Thus this gives a way to locate the
    transition point by $V_c=\exp(-D_{L=\infty}/C_{L=\infty})$, provided the two 
    parameters can be accurately determined.  Now we make a crude approximation
    that  taking $C$ and $D$ as constants and simply using the results at $L=18$
    as their values, namely $C=0.37$ and $D=-0.67$. Then we plot the $u/L$-th
    root of $\tau_1$ versus $V$ for each $L$ in Fig. \ref{fig:data:collapse}.
    The near-collapse of the data for each $L$ indicates that Eq.  \eqref{eq:tau1:LV} is
    plausible. 

\section{slow bound states under long-range interactions} \label{sec:bound:state}

\subsection{content of bound states} \label{sec:bound:state:content}

    For quantum integrable systems, existence of bound states as
    quasi-particles is well established \cite{bethe1931zur,
    takahashi2005thermodynamics}.  The success of GHD just relies on identifying
    those quasi-particles as charge carriers.    
    In particular, for the XXZ model a bound
    state is referred to as an $n$-string, which corresponds to a sequence of
    $n$ flipped spins (with the 1-string reduced to a single magnon).  It has a
    group velocity $ \sim \Delta^{-(n-1)}$ \cite{woellert2012solitary} and
    scatters forwardly with one another.  Note that at large $\Delta$  and $n$,
    its velocity is so slow that it resembles a contiguous block of  $n$
    localized spins \cite{mossel2010relaxation}.

    \begin{figure}   
        \scalebox{0.55}[0.55]{\includegraphics{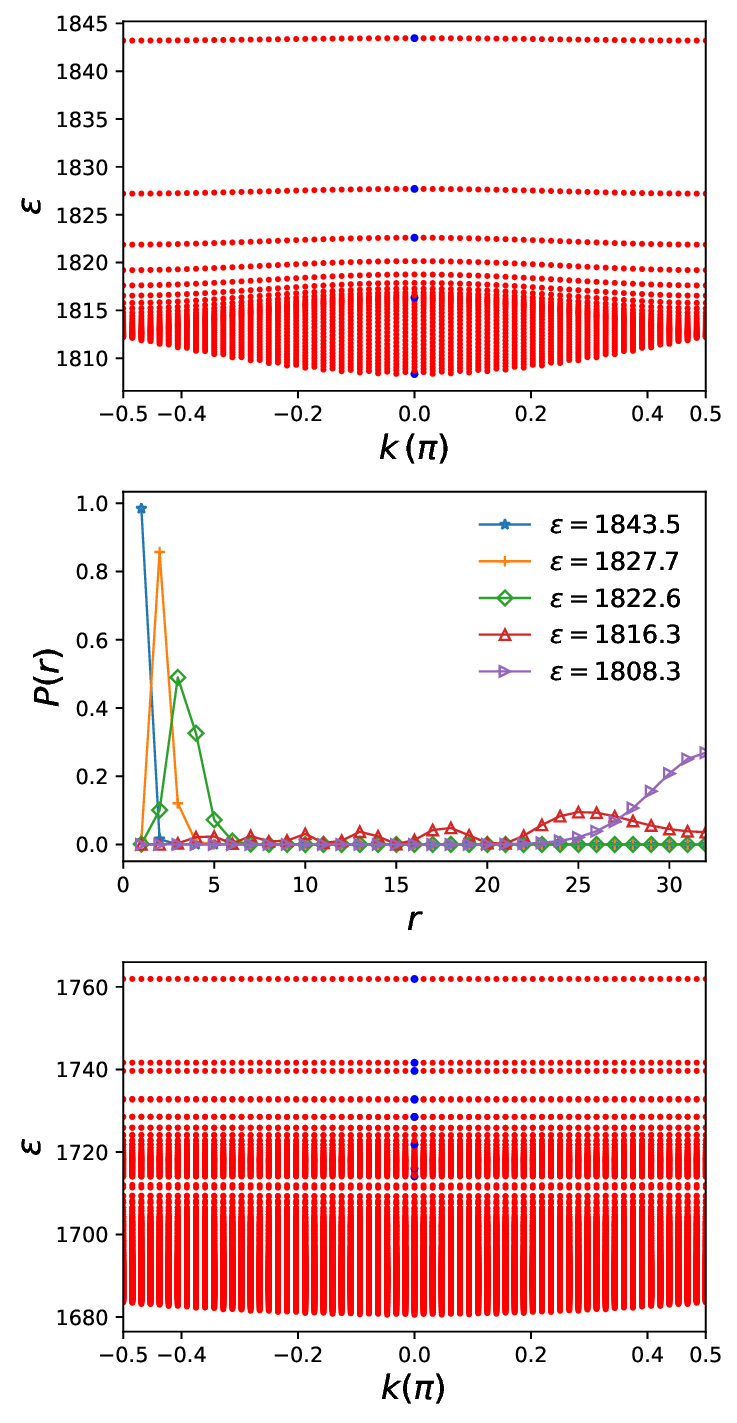}}
        \caption{\label{fig:bound:state} 
        Top panel: 2-particle spectrum of the Hamiltonian \eqref{eq:ham}  at
        $V=32$ on a lattice of size $L=64$.  
        The horizontal axis represents the momentum $k$ of the states; the
        vertical axis represents the energy $\varepsilon$. 
        Middle panel: distributions of two-particle separations $P(r)$ for five
        states in the $k=0$  sector, whose energies are shown in the legend and
        marked by blue dots in the top panel. 
        Bottom panel: 3-particle spectrum for the same $V$ and $L$ as
        the top panel.  
        }
          \end{figure}
    
    For generic quantum lattice models, bound states should also exist, which
    does \emph{not} rely on integrability, but on a limited band width. We
    expect that they should also play an essential role in transport, especially
    in the strong coupling regime.  For $n=2$ particles, existence of bound
    states can be proven for general interaction potentials, no matter they are
    attractive or repulsive
    \cite{kagan1984localization,winkler2006repulsively,valiente2010lattice}.
    For $n>2$ particles, there still lacks a general theory
    \cite{kagan1984localization,mattis1986few, valiente2010three,
    kornilovitch2013stability}.  However, existence of them may be anticipated
    from a simple energy conservation perspective: when the potential energy of
    a compact $n$-particle cluster is much greater than $n$ times the band
    width, it can't decay into spatially far-separated smaller pieces. These
    arguments should also hold for lattice models with long-range interactions
    \cite{valiente2010lattice}.  In the following, we show direct evidence for
    this for the Hamiltonian \eqref{eq:ham}, by numerically diagonalizing it for
    a system of $n$ particles on a ring lattice, for only small $n$'s.   
    
    First the top panel of Fig. \ref{fig:bound:state} shows the 2-particle spectrum  
    on a lattice with $L=64$ sites at a coupling strength $V=32$.  The few
    branches of energy bands on the top are bound states, while the continuum of
    states beneath are scattering states. To prove this, we measure the 
    two-point correlation functions $C^{(2)}(i,j)= \langle \psi |{{\hat
    n_i}{\hat n_{j}}} |\psi \rangle$ for the eigenstates $|\psi\rangle's$, based on which the
    probability of finding  the two fermions with a distance $r$ is
    $P(r)=\sum_{i=1}^{L}{C^{(2)}(i,i+r)}= L C^{(2)}(1,1+r)$. The second equality
    holds because of translation invariance. Note that all
    possible different $r$ values are $1,2,\dots,L/2$ on the ring lattice and 
    $\sum_{r=1}^{r=L/2}{P(r)}=1$.  Under these definitions, an eigenstate
    $|\psi\rangle$ should be a bound state if $P(r)$ is none-zero for only
    relatively  small $r$, otherwise, it is a scattering state. 
    Here we show $P(r)$ in the middle panel of Fig. \ref{fig:bound:state} for
    only five representative states, which are all in the zero momentum sector
    and indicated by blue dots in the top panel.  
    For the first three states  in the top bands, $P(r)$
    has pronounced peaks at $r=1,2$ and $3$, respectively, while being depressed for 
    large $r$. This shows that they
    are all bound states, and that the main contributions in the two-particle
    configurations are, respectively,  ``11'', ``101'' and ``1001'' (here 1
    represents an occupied site, 0 for an empty site, and trailing 0's are
    omitted for clarity). In contrast, for the two states in the continuum
    region, $P(r)$ is none-zero for a wide range of $r$, so they are scattering
    states.

    \begin{table}[h]  
    \centering  
    \caption{
        An (incomplete) list of 3-particle bound states in the zero momentum
        sector. The four columns from left to right are: state numbers (numbered from highest
        to lowest energy in that sector), energies ($\varepsilon$) of the states,
        strings ($s$) representing particle configurations and the probabilities 
        for each configurations $P(s)$, respectively. For each state, only
        configurations whose probability is larger than 0.4 are listed. 
        }  
    \begin{small}
        \begin{tabular}{p{2.1cm}p{1.5cm}p{1.5cm}p{1.2cm}}
        \toprule  
        state number     & $\varepsilon$  & $s$   & $P(s)$ \\
        \hline
        \hline
        \#1 & $1761.92$ 	 & 111 & 0.996 \\
        \hline
        \#2 & $1741.61$ 	 & 1011 & 0.494 \\
             && 1101 & 0.494 \\
        \hline
        \#3 & $1739.64$ 	 & 1101 & 0.487 \\
             && 1011 & 0.487 \\
        \hline
        \#4 & $1732.83$ 	 & 11001 & 0.461 \\
             && 10011 & 0.461 \\
        \hline
        \#5 & $1732.66$ 	 & 10011 & 0.469 \\
             && 11001 & 0.469 \\
        \hline
        \#6 & $1728.49$ 	 & 110001 & 0.416 \\
             && 100011 & 0.416 \\
        \hline
        \#7 & $1728.49$ 	 & 100011 & 0.418 \\
             && 110001 & 0.418 \\
        \hline
        \#14 & $1721.82$ 	 & 10101 & 0.941 \\
        \hline
        \#52 & $1716.00$ 	 & 101001 & 0.470 \\
             && 100101 & 0.470 \\
        \hline
        \#64 & $1714.07$ 	 & 100101 & 0.423 \\
             && 101001 & 0.423 \\
        \bottomrule  
        
    \end{tabular}
    \label{tab:three:particle}
    \end{small}
    \end{table}
    
    Next we consider the system with three fermions. Three-particle states are
    more cumbersome to characterize, for which the three-point correlation
    functions $C^{(3)}(i,j,k) = \langle \psi |{\hat n_i}{\hat n_j}{\hat
    n_k}|\psi \rangle$ are needed. 
    Besides that, we introduce some notations and terminologies for better
    describing these states.  We still use strings made up of 1's and 0's to
    denote particle configurations (up to translations on the ring lattice): ``111''
    means they occupy three contiguous sites; ``1101'' means two of them are
    nearest-neighbored and the other one is to the right of them but separated
    by 1 site, and so forth.  Then the probability $P(s)$ for a configuration
    $s$ found in a state $|\psi\rangle$ is determined by the three-point
    function, for example, $P(\text{``111''})=L C^{(3)}(1,2,3)$ and
    $P(\text{``1101''})= L C^{(3)}(1,2,4)$.  
    A configuration with a significant probability will be called as a primary
    configuration. Two configurations which are energetically equivalent, e.g.
    ``1101'' and ``1011'',  are called resonant configurations
    \cite{schiulaz2015dynamics, roeck2014asymptotic}. 
    
    The three-fermion spectrum is shown in the bottom panel of
    Fig. \ref{fig:bound:state}, where the system parameters are the same as the
    case of $n=2$.  It suffices to consider only the zero momentum sector,
    because the contents of bound states in different sectors will be similar.
    There are  561 states in all in that sector. For each state, we have
    measured the probability of occurrence of every configuration.  Table
    \ref{tab:three:particle} lists only the primary configurations for certain
    states, which are marked by blue dots in the bottom panel: For the 1st state
    at $\varepsilon=1761.92$, there is only one primary configuration, with its
    probability equal to 0.996. For the 2nd state at $\varepsilon=1741.61$,
    there are two primary configurations ``1101'' and ``1011'', which are in
    resonance, and their probabilities are both equal to 0.494. The 3rd state at
    $\varepsilon=1739.63$ is nearly degenerate with the 2nd state. It has the
    same primary configurations as the former, only that the probability goes
    down slightly to 0.487.  As a matter of fact, the main difference between
    the 2nd state and the 3rd state is that  the former is of odd parity while
    the latter is of even parity.  Likewise, the 4th and 5th states are nearly
    degenerate and have the same primary configurations, but differ in parities.
    It is easy to see that these states are all bound states, so should be the
    rest states in the table.

    \begin{figure}   
        \scalebox{0.55}[0.55]{\includegraphics{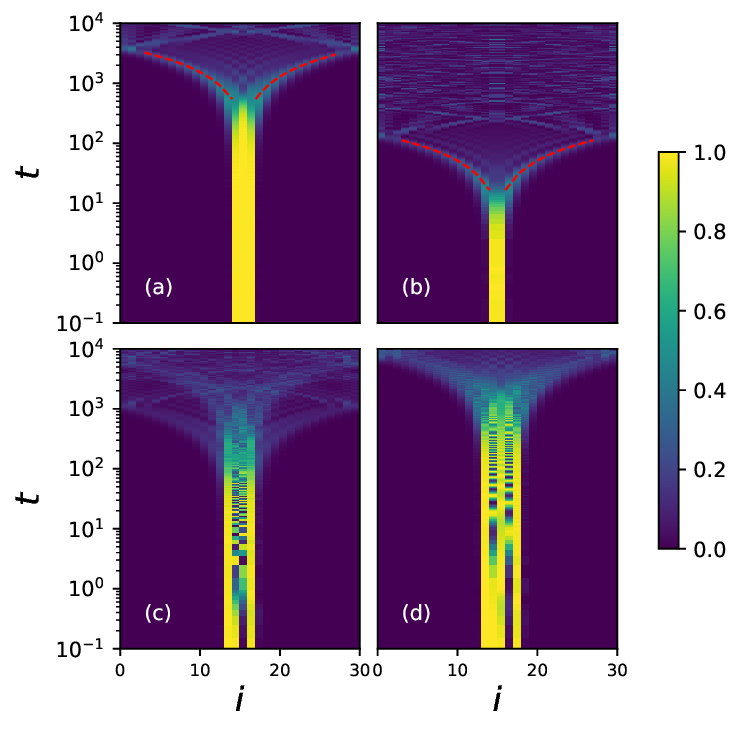}}
        \caption{
        \label{fig:bound:state:velocity}
        Time evolution of localized blocks of particles.  The initial states are
        product states $|0\dots0 s 0\dots 0\rangle$, where $s$ for (a)-(d) are
        ``111'', ``11'', ``1101'' and ``11101'', respectively. The time
        evolution is obtained by using ED with PBC; the system sizes are all $L=30$
        and the coupling strengths $V=32$; the color map encodes densities of
        particles.  In (a) and (b), the dashed lines are fittings of the peak
        positions of the densities in the wave fronts to a linear function
        $i=\pm  v_\text{g,max} t  + const.$, with \vgm being 1/247.3 and 1/8.7
        respectively.
        }
        \end{figure}
    
    From the above results we see that not only do bound states exist under
    long-range interactions, but their types are much richer than that of the
    short-ranged XXZ model. We will not study the general cases of
    $n>3$, but one may appreciate that at large couplings for {\emph {every}}
    compact particle configuration (up to resonances) there should be a
    corresponding bound state.  We also note that in an $n$-particle spectrum
    the high energy states are bound states, while the states in the bottom are
    scattering states (in the middle the states hybridize bound states and
    scattering states). The high energy bound states move slowly, while the latter
    move fast. Details of this point are explained in the next subsection.
    

\subsection{velocities of bound states, and duality between 
    bound states and localized particle blocks} \label{sec:bound:state:velocity}

    The single-particle states (singletons, or magnons in the language of spins) have
    the dispersion $\varepsilon(k)=-2 \cos(k)$.  Therefore the maximal value of
    their group velocities, denoted by $v_\text{g,max}$, equals $2$, which does
    not depend on the coupling strength. 
    For bound states with $n\geq 2$, their maximal group velocities loosely
    speaking behave like 
    \begin{equation}
        \label{eq:state:velocity}
        v_\text{g,max} \sim V^{-(n-1)}
    \end{equation}
    in $n$-th order of perturbation theory, so they can be very slow at large
    $V$ or $n$. And $n$ can be seen as an effective mass of a quasi-particle.  A
    more concrete and precise form of \vgm should nevertheless depend also on
    the primary configurations of the bound states.  In particular, when a primary
    configuration is resonant, there can be internal dynamics, which will be
    made clearer in the following discussions. 
    
    At large couplings, there is one kind of duality between $n$-particle bound
    states and localized $n$-particle blocks when they have corresponding
    configurations.  By duality we mean that they are connected approximately by
    Fourier transformation, as they are respectively eigenstates of momentum
    and position operators, and the connection is sharpened
    with increasing $V$.  This can be seen as a generalisation for that of the
    XXZ model \cite{vlijm2015quasi,woellert2012solitary,mossel2010relaxation}.  Based on
    the duality one may visualize the motion of the bound states by looking at 
    evolution of corresponding localized blocks. Here time evolution of
    four simplest configurations at $V=32$ is illustrated in
    Fig. \ref{fig:bound:state:velocity}.  Each of the four blocks delocalizes
    with time due to moving of its dual bound states and  other states
    (contributions from the other states are however negligibly small at large $V$).
    The wave fronts of the fastest modes form linear light cones, which can be
    clearly seen for the two non-resonant configurations ``111''  and ``11''
    (see subplots (a) and (b)).    The \vgm of them can be determined by
    measuring the slopes of the light cones, which are 1/247.3 and 1/8.7
    respectively, differing by a factor about $V$ and in agreement with
    Eq. \eqref{eq:state:velocity}.  The other two configurations ``1101'' and
    ``11101'' are both resonant, whose time evolution is a bit more
    complicated  (see subplots (c) and (d)).  As a whole, they also move slowly,
    but they can contain faster internal dynamics. The  two resonant
    configurations are of first and second order, respectively. Here the order
    $p$ of a resonant configuration is defined as the number of displacements
    required to change it to its resonant counterpart \cite{schiulaz2015dynamics}. Then
    quantitatively the speed of the internal particles in it should scale as
    $\sim V^{-(p-1)}$. In particular, the central particle in the first-order
    resonance has a velocity $\approx 1.5$, which is fast and does not depend on
    $V$.

    Note that the duality approximately holds for only large $V$. When $V$ is
    reduced, a localized $n$-particle block will receive more and more
    contributions from lighter and faster $m$-particle states, for all $m<n$.
    This is similar to the results of the XXZ model \cite{vlijm2015quasi}, and
    we will not show numerical evidence for this for the present model. So
    delocalization of a block of localized particles is quickened by a smaller
    $V$ for dual reasons: it is decomposed more into lighter types of
    quasi-particles, and the velocities of each type of quasi-particles scale 
    faster (through Eq. \eqref{eq:state:velocity}). This point is crucial
    for understanding the result in the last section that a thermalization to
    quasi-MBL transition occurs when varying $V$.

\comment{
    \subsection{collision properties} \label{sec:bound:state:collision}
        Transport and relaxation not only depend on the content of quasi-particles
        but also on their collision properties, the latter being more challenging to
        be determined. We only pursue one aspect of that in an indirect way.
        Specifically, we collide a Gaussian wave packet of a single particle (or  a
        magnon, in the language of spins, see appendix \ref{sec:app:A} for details
        of its definition) with several initially localized cluster of particles (or
        spin blocks) \cite{ganahl2013quantum,vlijm2015quasi}, as plotted in
        Fig. \ref{fig:bowling:lr}. For each case the block remains stable for a long
        time, thus close to an eigenstate of the Hamiltonian.  Nervelessly, these
        blocks are themselves not the bound states, but they do have very close
        relations. It could be shown that the former have a large overlap with the
        latter (upon superposition with different momenta), if they correspond to the
        same particle configuration 
        \footnote{
        One may make an analysis similar to that made in \cite{mossel2010relaxation}
        to justify this point. Alternatively, one may understand it by the following
        arguments: A bound state and a corresponding classical configuration of
        particle cluster is in a sense dual to each other, related by Fourier
        transformations. The former is largely a superposition of the latter up to
        translations, while the latter is largely a superposition of the former for
        all momenta. This argument may be oversimplified for certain cases, where a
        bound state can has two (or even more) primary configurations.  }.  So the
        collision properties between 1-particle states and the bound states can be
        inferred thereof. From subplot (b), we can see that the single particle is
        (almost) always backscattered by the cluster (superposition of bound states)
        under the Coulomb potential. The backscattering should not altered by the
        fine details of a remote bound state, as exemplified by subplot (c) and (d).
        It is in stark contrast with the short-range limit of the model, where there
        is only forward scattering due to integrability [see subplot (a)]. 

    }    

\subsection{interpretations of the relaxation processes} \label{sec:bound:state:explain}
    
    \begin{figure}   
        \scalebox{0.55}[0.55]{\includegraphics{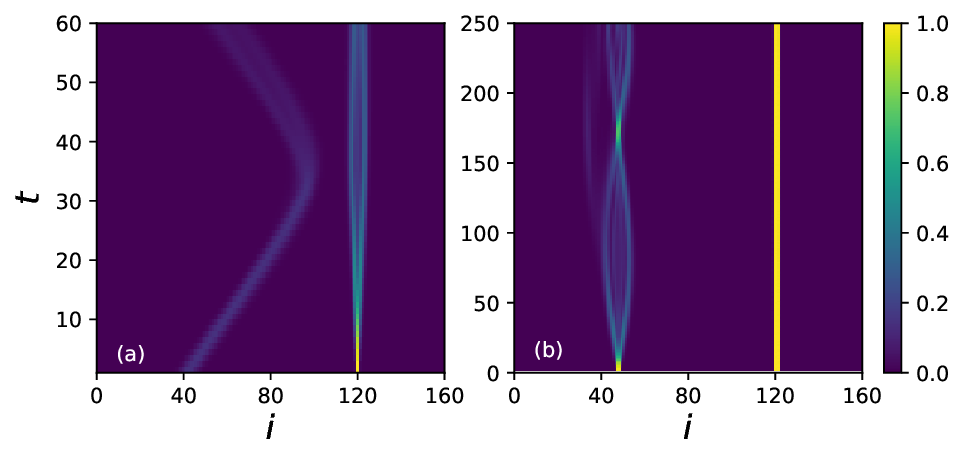}}
        \caption{
        \label{fig:bowling:lr} 
        Collisions between quasi-particle wave packets: Panel (a) depicts a Gaussian wave
        packet of a singleton colliding with 2-particle wave packets, the latter being
        decomposed from a localized 2-particle block. Panel (b) shows evolution
        of two initially localized blocks ``11'' and ``111'' with a separation
        of 70 lattice sites.  The dynamics are obtained by using TDVP for $V=32$,
        $L=160$ and PBC.  The color map encodes densities of particles.
        }
        \end{figure}

    At a large coupling  the only fast modes are the singletons and the
    first-order resonant processes, while other modes are all slow and differ in
    orders of magnitude of $V$.  Given existence of slow quasi-particles, to
    account for the macroscopic transport and relaxation processes, one still
    needs to know how these quasi-particles interact with one another, which we 
    discuss next. The discussions are first restricted to the large coupling
    regime, where the physical picture is simpler and the  above-stated duality
    can be utilized.  Depending on the density of particles on the lattice, the
    physical pictures can be very different.
   
    For very low particle densities the physical picture is this: far apart
    quasi-particles are moving on the lattice, and faster ones are jammed by
    slower ones. We illustrate this point by two examples of few-body dynamics.
    The first example is a right-moving Gaussian wave packet of a singleton
    colliding with a  2-particle wave packet, the latter being decomposed from a
    localized 2-particle block (the detailed definition of the initial state  is
    given in the appendix).  It turns out that they are
    backscattered before approaching very close to each other, as shown in
    Fig. \ref{fig:bowling:lr}(a).  This is in stark contrast with the XXZ model
    \cite{ganahl2013quantum,vlijm2015quasi}, where the nearest-neighbor
    interactions lead to only forward scatterings. The second example is a
    2-particle block interacting with a 3-particle block. The quasi-particles
    decomposed from the 2-particle block are also backscattered by the more
    stable 3-particle block, so that the motion of the former is constrained
    (see Fig. \ref{fig:bowling:lr}(b)).  From these two simple examples,  we
    infer that two quasi-particles of general types may be always backscattered
    by each other under the long-range Coulomb potentials, provided they are
    initially far apart.  
    We will however not delve deeper for the low particle densities, as the
    macroscopic relaxation processes presented in section \ref{sec:results:f}
    are at half filling, which is discussed next.

    \begin{figure}   
        \scalebox{0.55}[0.55]{\includegraphics{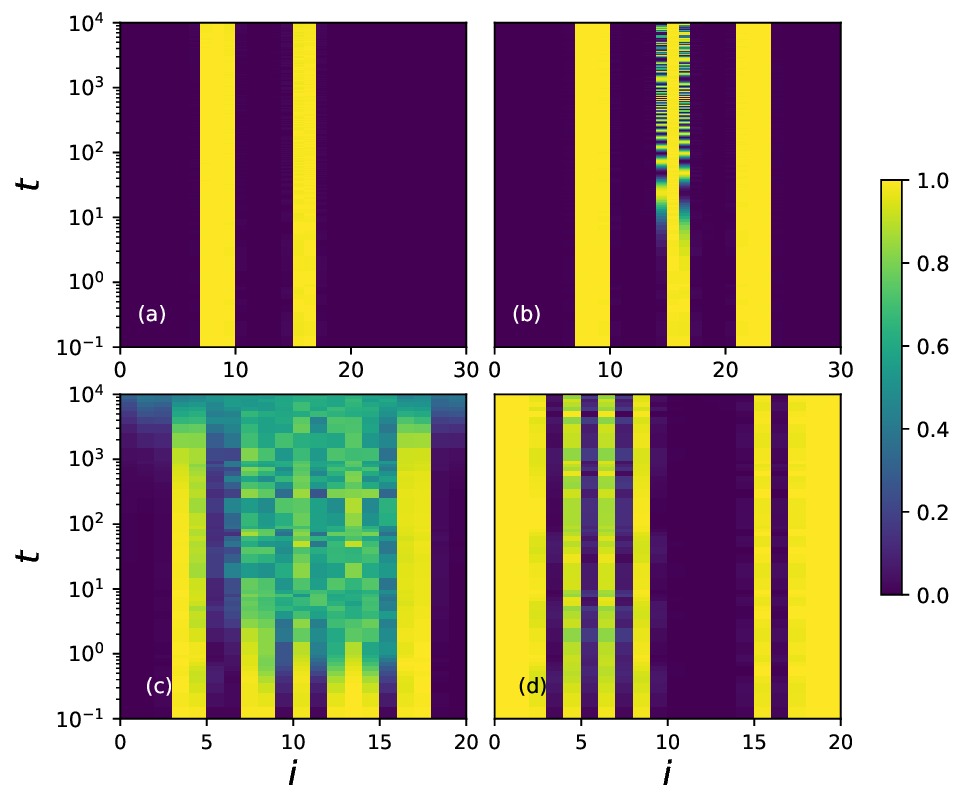}}
        \caption{
        \label{fig:cluster} 
        Time evolution for certain product state initial states: For subplots
        (a) and (b) the particle configurations are ``1110000011'' and
        ``11100000110000111'', respectively, both on a lattice with 30 sites. 
         For (c) and (d) both are 10 particles on a lattice with 20
        sites, where the initial particle configurations can be read off from the
        graphs. The color map encodes densities of particles. The dynamics are
        obtained by using ED with PBC; the coupling strength $V=32$ for all
        cases. 
        }
        \end{figure}

    At or close to half filling, the crowdedness of the particles leads to two
    competing effects. On the one hand, it reinforces the stability of small
    particle blocks and localizes the particles. For example when two localized
    blocks ``111'' and ``11'' are placed nearby, say five sites away, the
    stability of them are both reinforced (see Fig. \ref{fig:cluster}(a)).  This
    can potentially lead to a large and stable cluster, when more particles are
    added nearby.  But on the other hand, the crowdedness also leads to numerous
    resonant configurations, that tend to delocalize certain particles.  For
    example if another ``111'' block is added to four sites to the right of the
    previous 5-particle system, then the two fermions in the middle move faster
    due to resonance (see Fig. \ref{fig:cluster}(b)).  We note that the
    former effect dominates at large length scales, whereby large and stable
    clusters may form; while the latter is constrained to be in small length
    scales, inside the clusters; but eventually the clusters thermalize locally
    through the resonances. 
    
    The time scales for local thermalization of the clusters vary significantly,
    which depend on specific configurations.  For example comparing the two
    configurations of Fig. \ref{fig:cluster}(c) and (d), both being  a cluster of
    10 particles on a lattice with 20 sites, the former thermalizes faster than
    the latter.  Usually, for a given coupling strength, the clusters with high
    energy densities (i.e. containing long contiguously occupied sites) or
    containing resonances at only  high orders thermalize slower. Now imagine a
    system with more particles on a larger lattice than the 
    examples of (c) and (d). Then in some regions the clusters will thermalize fast
    and in some other regions they do so slowly. While the point is that the
    motion of the thermalized (or ``delocalized'') regions is still constrained
    by surrounding more stable clusters, which prevents the entire system from
    thermalizing. In other words, local thermalization can be embedded in
    global quasi-localization.  One may continue this thought and consider the
    system just stated to be on an even larger lattice, and on and on.  These
    descriptions would in the end lead to the picture of a hierarchy of
    stable clusters of particles on many different length scales.

    Each length scale $\ell$ of the stable clusters determines a local thermalization
    time scale.  While the most important is the one with the maximal size
    $\ell_{max}$, which determines the relaxation time  of the entire system.
    For a system described by an ensemble at a certain high temperature, the
    relevant quantity is an ensemble-averaged value $\langle\ell_{max}\rangle$. 
    We expect that when $V$ is large, this value should be proportional
    to the system length, $ \langle \ell_{max}\rangle \propto L$, so that
    this together with Eq.  \eqref{eq:state:velocity} is roughly in accordance
    with the exponential scaling of $\tau_1$ with $L$ and the power-law scaling
    of it with $V$ (i.e.  Eqs.  \eqref{eq:tau1:L} and \eqref{eq:tau1:V}).  As
    $V$ decreases, the clusters of particles are less stable due to the reasons
    stated in the final paragraph of the last subsection. Then we expect that
    when $V$ is smaller than some threshold, $\langle \ell_{max} \rangle $
    should saturate as $L$ increases, and the stable clusters are all relatively
    small-sized.  So these arguments provide a microscopic  mechanism for the
    quasi-MBL to thermalization transition.


    No matter the average maximal size of the stable clusters grows linearly
    with $L$ or not, for any finite $L$, clusters on all length scales will
    gradually delocalize, starting from the lowest-order resonances.  The
    intermediate time scales in the transport and relaxation processes are
    related to different sizes of clusters (quasi-particles) and different
    orders of resonances. 
    Specifically, the fast decay of $f$ for $t\leq 1$ (i.e. stage (i)) is
    completely due to motion of the first-order resonances and the
    singletons. The velocities of these fast modes do not depend on $V$, but the
    densities of them do, that is why $f$ drops to lower values for smaller $V$.
    These facts are also consistent with $z\approx 1$ at $t=1$. The transient
    periods (stages (ii)) for both $f(t)$ and $z(t)$ should be because of
    further relaxations related to these fast modes. The slow change of $f$ and
    $z$ with time in stages (iii) should be caused by successive relaxation of each 
    intermediate-sized clusters, through each higher order resonances. However, a
    quantitative explanation of why they are approximately in logarithmic forms
    needs further investigation.

\section{conclusion}  \label{sec:conclusion}

    We studied  transport and relaxation of the fermion model with long-range
    Coulomb interactions for a wide range of couplings.  By extracting two
    time-dependent quantities $z(t)$ and $f(t)$ from out-of-equilibrium
    dynamics, we showed that when tuning the coupling strength of the long-range
    interactions, there is a dynamical phase transition at high temperatures.
    For large couplings, the system exhibits anomalous subdiffusive transport
    (through the behavior of $z$) and at the same time quasi-localization
    (through $f$), whereby a correspondence between the two descriptions is established. 
    For intermediate couplings, the system exhibits normal diffusive transport
    and thermalization after certain time scales.  
    However, even in this ``normal'' regime, both $z$ and $f$ change slowly with
    time before reaching those time scales, which can be fitted by logarithmic
    functions. 
    This shows that the usual assumption of rapid local
    chaotic thermalization of Hydrodynamics is false for the present
    non-integrable model.
    

    We have tried to interpret the macroscopic transport and relaxation
    processes by studying few-particle problems on the lattice. We found that
    there is a richness of types of bound states under the long-range Coulomb
    force. And the motion of all quasi-particles all slow down except the
    singletons and first-order resonances, when the coupling strength increases.
    Besides, for many particles at large densities, the long-range interactions tend to
    bind localized blocks together to form large clusters, but at the same
    time, they also lead to various internal resonant processes. In the end there
    should be a hierarchy of clusters on different length scales. 
    We argue that at large couplings there should be giant immobile clusters,
    which gives an interpretation of the structure of the quasi-MBL states.

    
    
    Every quantum lattice model, being it integrable or not, should be able to
    produce bound states, and the bound states are slow moving. But not every
    model supports slow transport at large couplings and high temperatures.
    Another decisive factor yet required is formation of large and stable clusters
    of particles. This depends on specific forms of interactions.  It
    appears that long-range power-law interactions usually suffice for this
    requirement, since where slow relaxation dynamics are found in the present model
    and in previous works \cite{kagan1984localization, schiulaz2015dynamics,
    spielman2022slow}. Nevertheless, we expect that similar slow transport may
    be found in a much wider range of models. It is interesting to determine
    the minimal conditions for the slow dynamics in future works.

\appendix*
\section{initial state for the collision dynamics}  \label{sec:app:A}

Following Refs. \cite{ganahl2013quantum} and \cite{ulbricht2009is}, the initial state
$|\psi(0)\rangle$ of the dynamics is created by acting the operator (up to
normalization) 
\begin{eqnarray}
      \sum\limits_x {\exp } \left( { - \frac{{{{\left( {x - {x_0}}
      \right)}^2}}}{{2{\sigma ^2}}}} \right)\exp \left( {i\left( {x - {x_0}}
      \right){k_0}} \right)c_x^\dag 
    \end{eqnarray}
on a product state $|0\dots0110\dots0\rangle$ for a block of two localized particles.
This operator creates a right-going Gaussian wave packet with momentum
$k_0=-\pi/2$, width $\sigma=4$, and center position $x_0$ as depicted in the
figure. 

\comment{ 
 
    \section{three particle states}  \label{sec:app:B}



    \begin{longtable}{lll}
        \caption{Probability of certain particle configurations of the states  in
        the zero momentum sector for a 3 particles system. The system  length $L=64$
        and the coupling strength $V=32$. There are 651 states in this sector, which
        are numbered  from highest to lowest energy. In each group of data shows the
        the numbering of a state with its energy $\varepsilon$ (the first line), and . all
        some particle configurations with their probabilities (the rest lines).
        Only configurations whose probability is larger than 0.1 are listed. 
        }  \\
        
        \hline
        \#1 & $\varepsilon=1761.92$ & \\
             & 111 & 1.00 \\
        \hline
        \#2 & $\varepsilon=1741.61$ & \\
             & 1011 & 0.49 \\
             & 1101 & 0.49 \\
        \hline
        \#3 & $\varepsilon=1739.64$ & \\
             & 1101 & 0.49 \\
             & 1011 & 0.49 \\
        \hline
        \#4 & $\varepsilon=1732.83$ & \\
             & 11001 & 0.46 \\
             & 10011 & 0.46 \\
        \hline
        \#5 & $\varepsilon=1732.66$ & \\
             & 10011 & 0.47 \\
             & 11001 & 0.47 \\
        \hline
        \#6 & $\varepsilon=1728.49$ & \\
             & 110001 & 0.42 \\
             & 100011 & 0.42 \\
        \hline
        \#7 & $\varepsilon=1728.49$ & \\
             & 100011 & 0.42 \\
             & 110001 & 0.42 \\
        \hline
        \#14 & $\varepsilon=1721.82$ & \\
             & 10101 & 0.94 \\
        \hline
        \#52 & $\varepsilon=1716.00$ & \\
             & 101001 & 0.47 \\
             & 100101 & 0.47 \\
        \hline
        \#64 & $\varepsilon=1714.07$ & \\
             & 100101 & 0.42 \\
             & 101001 & 0.42 \\
        \hline
        \#69 & $\varepsilon=1708.14$ & \\
             & 1001001 & 0.64 \\
        \hline
        \#82 & $\varepsilon=1703.89$ & \\
             & 100010001 & 0.43 \\
             & 100100001 & 0.11 \\
             & 100001001 & 0.11 \\


    \end{longtable}
    }

\section*{References}
%
    



\end{document}